\documentclass[reprint, amsmath,amssymb, aps]{revtex4-1}

\usepackage{graphicx}
\usepackage{dcolumn}
\usepackage{bm}
\graphicspath{{fig/}}
\usepackage[colorlinks=true,linkcolor=blue,citecolor=blue, urlcolor=blue]{hyperref}

\begin{document}


\title{Photoproduction of the $B_c$ meson at the future $e^+e^-$ colliders}

\author{Xi-Jie Zhan}
 	\email{zhanxj@cqu.edu.cn}

\author{Xing-Gang Wu}%
	\email{wuxg@cqu.edu.cn}

\author{Xu-Chang Zheng}
	\email{zhengxc@cqu.edu.cn}

\affiliation{Department of Physics, Chongqing Key Laboratory for Strongly Coupled Physics, Chongqing University, Chongqing 401331, P. R. China}

\date{\today}

\begin{abstract}

In the present paper, we study the photoproduction of various $S$-wave $B_c$ states, i.e. $B_c(1{}^1S_0)$, $B_c^*(1{}^1S_0)$, $B_c(2{}^1S_0)$ and $B^*_c(2{}^3S_1)$, at future $e^+e^-$ colliders within the framework of non-relativistic QCD. Two sources of the initial photons can be realized in the production, i.e., the LBS photon and the WWA photon. In addition to the direct photoproduction via the sub-process, $\gamma + \gamma \rightarrow B_c + b + \bar{c}$, we also calculate the resolved photoproduction via the sub-processes, $\gamma + g \rightarrow B_c + b + \bar{c}$ and $q + \bar{q} \rightarrow B_c + b + \bar{c}$ with $q=u,d,s,g$, respectively. Numerical results indicate that the contributions from the single resolved photoproduction are significant and even dominant at certain collision configuration. At the future high-energy and high-luminosity $e^+e^-$ colliders, the $B_c$ meson generated via the photoproduction mechanism is promisingly observable and can be well studied.
\end{abstract}

\maketitle

\section{\label{sec:1}Introduction}

The $B_{c}(1{}^1S_0)$ meson, including its excited states, are strong interaction systems composed of two heavy quarks with different flavors. This unique flavored bound-state system could be a good probe for studying the strong interaction among quarks and gluons, and the quantum chromodynamics (QCD). So far the $B_c$ mesons have been only observed at the hadronic colliders. Its ground state was discovered in 1998 by the CDF collaboration~\cite{CDF:1998ihx, CDF:1998axz}, and its excited $2S$ states were observed by ATLAS collaboration~\cite{ATLAS:2014lga} in 2014 and by CMS and LHCb collaborations~\cite{CMS:2019uhm, LHCb:2019bem} in 2019. Many theoretical studies have been done for $B_c$ production at hadron colliders~\cite{Chang:1992jb, Chang:1994aw, Kolodziej:1995nv, Berezhnoy:1994ba, Chang:1996jt, Berezhnoy:1996ks, Baranov:1997wy, Baranov:1997sg, Cheung:1999ir, Chang:2003cr, Chang:2004bh, Chang:2005bf, Chang:2005wd, Chang:2003cq, Chang:2005hq, Wang:2012ah, Bi:2016vbt, Chen:2018obq, Berezhnoy:2019yei, Zheng:2019egj}. A computer program, BCVEGPY, for the direct hadronic production of $B_c$ meson has been given in Refs.\cite{Chang:2003cq, Chang:2005hq, Wang:2012ah}. The next generation $e^+e^-$ colliders were proposed by some groups, such as the FCC-ee~\cite{FCC:2018evy}, the CEPC~\cite{CEPCStudyGroup:2018rmc, CEPCStudyGroup:2018ghi}, the ILC~\cite{ILC:2007bjz, Erler:2000jg} and so on. These future $e^+e^-$ colliders are planed to have the ability to run at several high collision energies with unprecedented luminosities. They are expected to be great platforms for many subjects including heavy quarkonium physics~\cite{Yang:2011ps, Chen:2013itc, Chen:2013mjb, Sun:2013liv, Chen:2014xka, Sun:2014kva, Sun:2015hhv, He:2019tig, Zhan:2020ugq, Zhan:2021dlu, Zhan:2022nck}.

As for the $B_c$ meson, there are two production modes at the $e^+e^-$ collider, i.e., the production via the $e^+e^-$ annihilation~\cite{Yang:2011ps, Yang:2013vba, Zheng:2015ixa, Berezhnoy:2016etd, Zheng:2017xgj, Zheng:2018fqv, Zhang:2021ypo, Yang:2022zpc} and the photoproduction mechanism~\cite{Chen:2014xka, Chen:2020dtu, Yang:2022yxb}. A computer program, BEEC, for the first production mode has been presented in Refs.\cite{Yang:2013vba, Yang:2022zpc}. As for the photoproduction mechanism, the $B_c$ meson can be produced via the photon-photon fusion such as $\gamma+\gamma \rightarrow B_c +b+\bar{c}$. Here  the initial photons can come from the bremsstrahlung effect, whose energy spectrum are well described in Weiz\"acker-Williams approximation (WWA)~\cite{Frixione:1993yw}; moreover, the laser back-scattering (LBS) can also provide high energy photon.
In addition to the above direct photoproduction, there are also channels of resolved photoproduction, where the photons participate in the hard process via their quark and gluon content~\cite{Klasen:2001cu}.
As a result, the process $e^+e^- \rightarrow e^+e^-B_c+X$ receives contributions from three channels, i.e., the direct, single-resolved and double-resolved photoproduction. All three channels should be considered in the calculation because they are of the same order in the perturbative expansion.
The resolved photoproduction of heavy quarkonium have been investigated in some literature~\cite{Klasen:2001cu,Li:2009zzu,Zhan:2020ugq,Zhan:2021dlu,Zhan:2022nck} and 
it is indicated that the single-resolved photoproduction can give significant or dominant contributions both in the color-singlet and color-octet channels.
While for the $B_c$ meson, only the direct photoproduction channel has been studied~\cite{Chen:2014xka, Chen:2020dtu, Yang:2022yxb} and it is worthy and reasonable to investigate the effects of the resolved channels according to the previous studies.

In this work, based on the framework of non-relativistic QCD (NRQCD)~\cite{Bodwin:1994jh}, we study the photoproduction of $B_c, B_c^*(1{}^3S_1), B_c(2{}^1S_0)$ and $B^*_c(2{}^3S_1)$ at future $e^+e^-$ colliders, considering two sources of the initial photons, i.e., the LBS photon and the WWA photon. In addition to the direct photoproduction $\gamma + \gamma \rightarrow B_c + b + \bar{c}$, we specially calculate the sub-processes of the resolved photoproduction, $\gamma + g \rightarrow B_c + b + \bar{c}$ and $q + \bar{q} \rightarrow B_c + b + \bar{c}$ with $q=u,d,s,g$, respectively. In Section~\ref{sec:2}, we present the formulation of our calculation. Numerical results and discussions are given in Section~\ref{sec:3} and a brief summary is in Section~\ref{sec:4}.

\section{\label{sec:2}Formulation}

\begin{figure}[htb]
	\includegraphics[width=.4\textwidth]{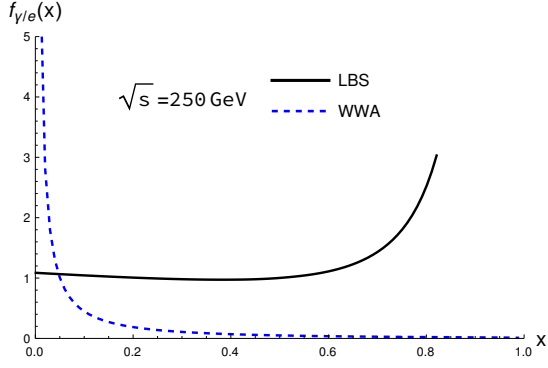}
	\caption{\label{fig:fre} The energy spectra of the LBS photon and the WWA photon.}
\end{figure}

\begin{figure*}
	\centering
	\includegraphics[width=.9\textwidth]{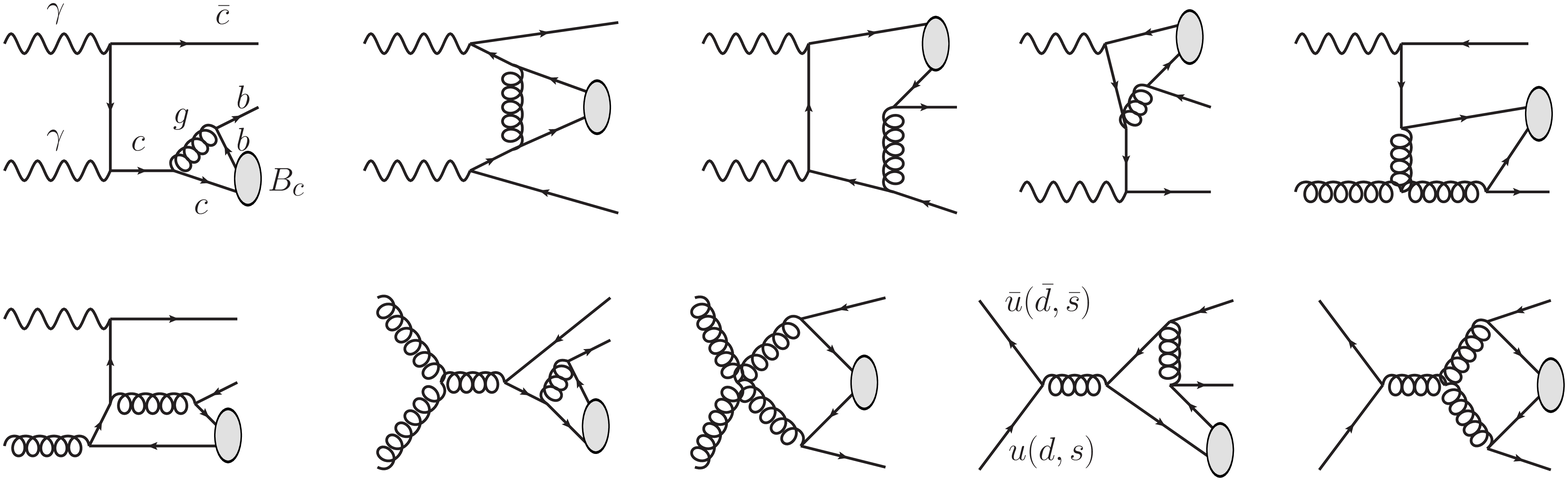}
	\caption{\label{fig:diag} Some typical Feynman diagrams for calculating the partonic cross section $\hat{\sigma}$ of $B_c$ photoproduction at $e^+e^-$ collider. The diagrams are drawn by JaxoDraw~\cite{Binosi:2003yf}.}
\end{figure*}

\begin{table}[b]
	\caption{\label{tab:cross-section-1}The integrated cross sections (in unit of fb) for the photoproduction of $B_c$ via the LBS photon and the WWA photon (in brackets), respectively. Three typical collision energies are taken and the cut $p_t>1$ is imposed. Three channels of Eq.~(\ref{eq:channel-1},\ref{eq:channel-2},\ref{eq:channel-3}) have been summed up.}
	\begin{ruledtabular}
		\begin{tabular}{cccc}
			$\sqrt{S}(\mathrm{GeV})$ & 250 & 500 & 1000\\
			\colrule
			$\sigma_{B_c}$ & 33.50(0.73) & 38.74(1.59) & 55.99(3.04)\\
			$\sigma_{B^*_c}$ & 175.06(7.14) & 177.49(14.09) & 240.69(24.82) \\
			$\sigma_{B_c(2{}^1S_0)}$ & 20.05(0.44) & 23.19(0.95) & 33.52(1.82)\\
			$\sigma_{B^*_c(2{}^3S_1)}$ & 104.80(4.27) & 106.26(8.43) & 144.09(14.86)\\
		\end{tabular}
	\end{ruledtabular}
\end{table}

\begin{table*}
	\caption{\label{tab:cross-section-2}
		The integrated cross sections (in unit of fb) of different channels of the photoproduction of $B_c$ via the LBS photon. Three typical collision energies, $250(500,1000)\mathrm{~GeV}$, are taken and the cut $p_t>1$ is imposed.}
	\begin{ruledtabular}
		\begin{tabular}{cccc}
			$channels$ & $\gamma+\gamma$ & $\gamma+g$ & $q+\bar{q}$\\
			\colrule
			$\sigma_{B_c}$ & $20.28(10.74,4.92)$ & $12.89(27.10,48.70)$ & $0.33(0.90,2.37)$ \\
			$\sigma_{B^*_c}$ & $114.90(54.93,23.42)$ & $59.03(120.05,211.15)$ & $1.14(2.51,6.12)$  \\
			$\sigma_{B_c(2{}^1S_0)}$ & $12.14(6.43,2.95)$ & $7.72(16.22,29.15)$ & $0.20(0.54,1.42)$  \\
			$\sigma_{B^*_c(2{}^3S_1)}$ & $68.78(32.88,14.02)$ & $35.34(71.87,126.41)$ & $0.68(1.50,3.66)$  \\
		\end{tabular}
	\end{ruledtabular}
\end{table*}

\begin{table*}
	\caption{\label{tab:cross-section-3}
		The integrated cross sections (in unit of fb) of different channels of the photoproduction of $B_c$ via the WWA photon. Three typical collision energies, $250(500,1000)\mathrm{~GeV}$, are taken and the cut $p_t>1$ is imposed.}
	\begin{ruledtabular}
		\begin{tabular}{cccc}
			$channels$ & $\gamma+\gamma$ & $\gamma+g$ & $q+\bar{q}$\\
			\colrule
			$\sigma_{B_c}$ & $0.68(1.33,2.37)$ & $0.05(0.19,0.58)$ & $0.001(0.005,0.017)$ \\
			$\sigma_{B^*_c}$ & $6.89(13.18,22.11)$ & $0.24(0.89,2.64)$ & $0.007(0.02,0.06)$  \\
			$\sigma_{B_c(2{}^1S_0)}$ & $0.41(0.83,1.46)$ & $0.03(0.11,0.35)$ & $0.0008(0.003,0.01)$  \\
			$\sigma_{B^*_c(2{}^3S_1)}$ & $4.13(7.89,13.24)$ & $0.14(0.53,1.58)$ & $0.004(0.013,0.036)$  \\
		\end{tabular}
	\end{ruledtabular}
\end{table*}

\begin{figure*}
	\includegraphics[width=.32\textwidth]{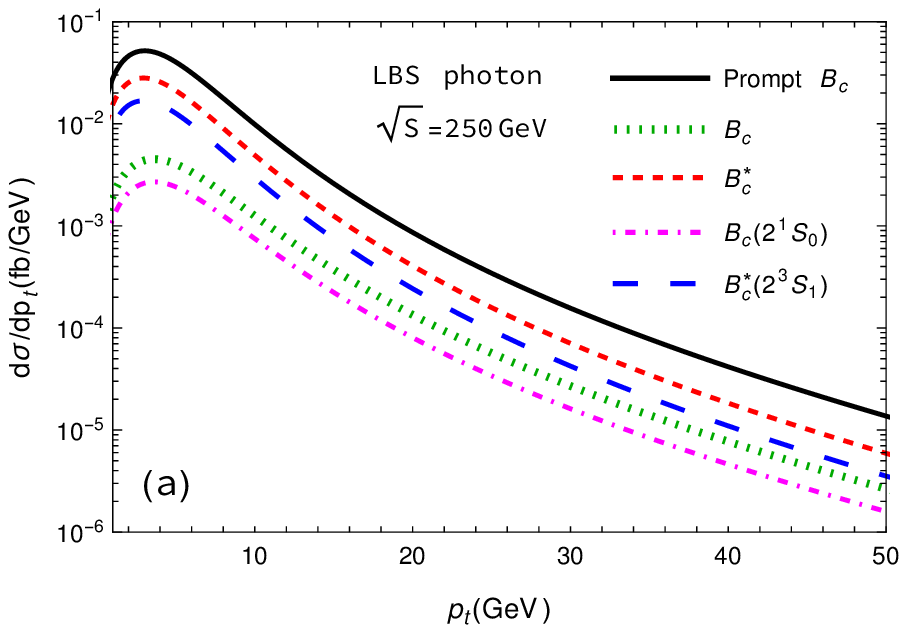}
	\includegraphics[width=.32\textwidth]{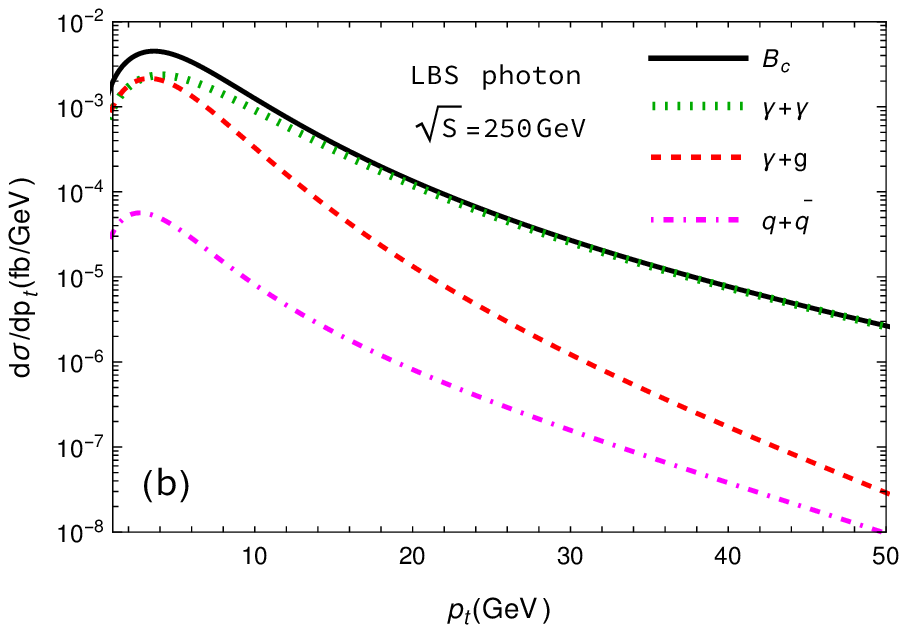}
	\includegraphics[width=.32\textwidth]{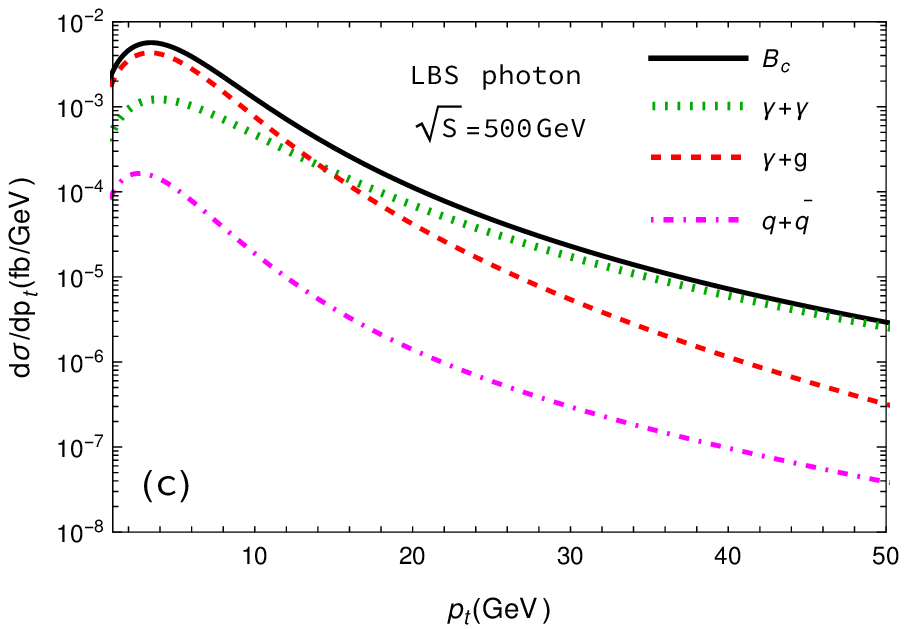}\\
	\includegraphics[width=.32\textwidth]{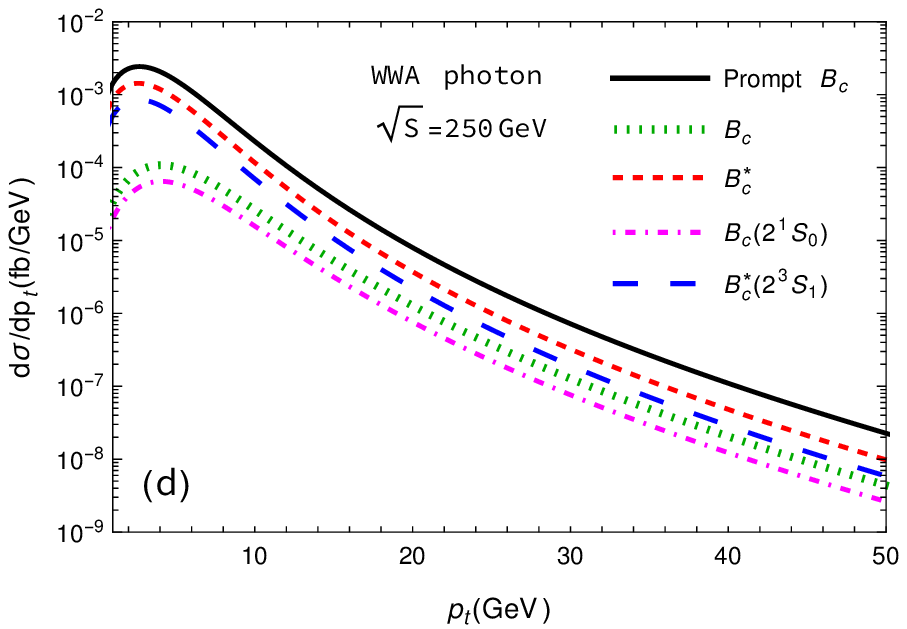}
	\includegraphics[width=.32\textwidth]{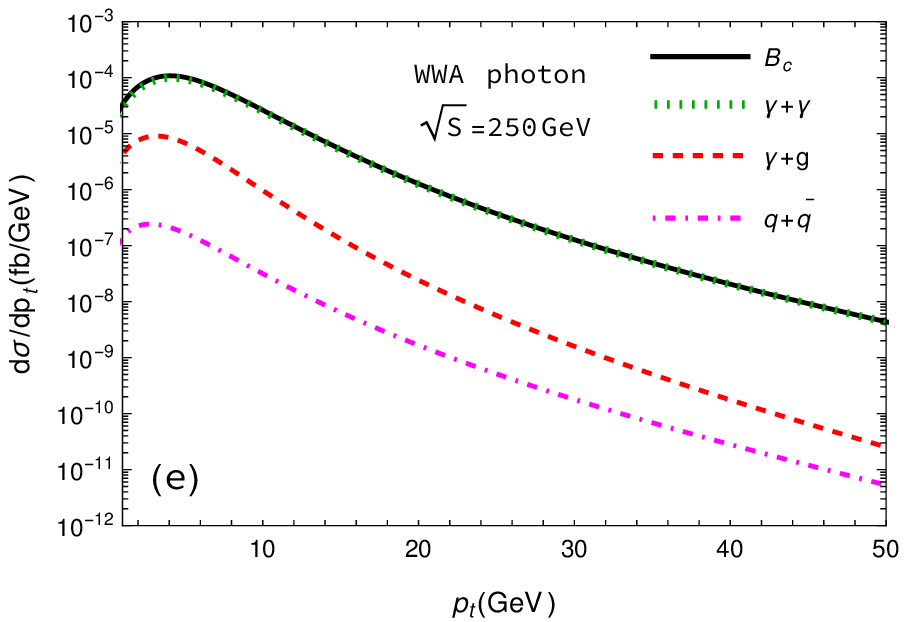}
	\includegraphics[width=.32\textwidth]{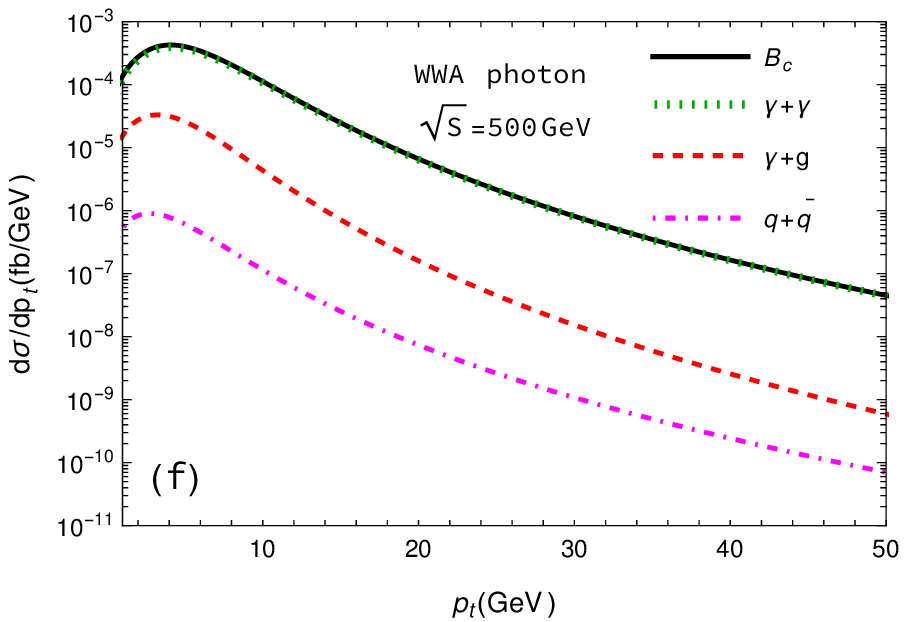}
	\caption{\label{fig:pt} The $p_t$ distributions for $B_c$ photoproduction. (a,d): $p_t$ distributions of four $B_c$ states and ``prompt $B_c$" means production of the ground $B_c$ after including the feed-down contributions from excited states with $100\%$ decay probability to it.  (b,c,e,f): $p_t$ distributions for the channels in Eq.~(\ref{eq:channel-1},\ref{eq:channel-2},\ref{eq:channel-3}) of the ground $B_c$ production.}
\end{figure*}

\begin{figure*}
	\includegraphics[width=.24\textwidth]{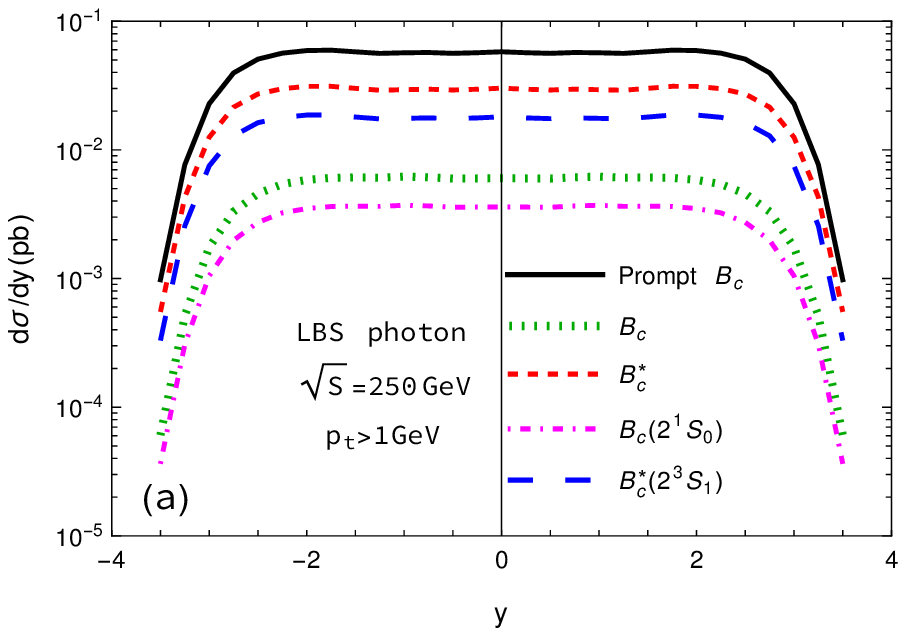}
	\includegraphics[width=.24\textwidth]{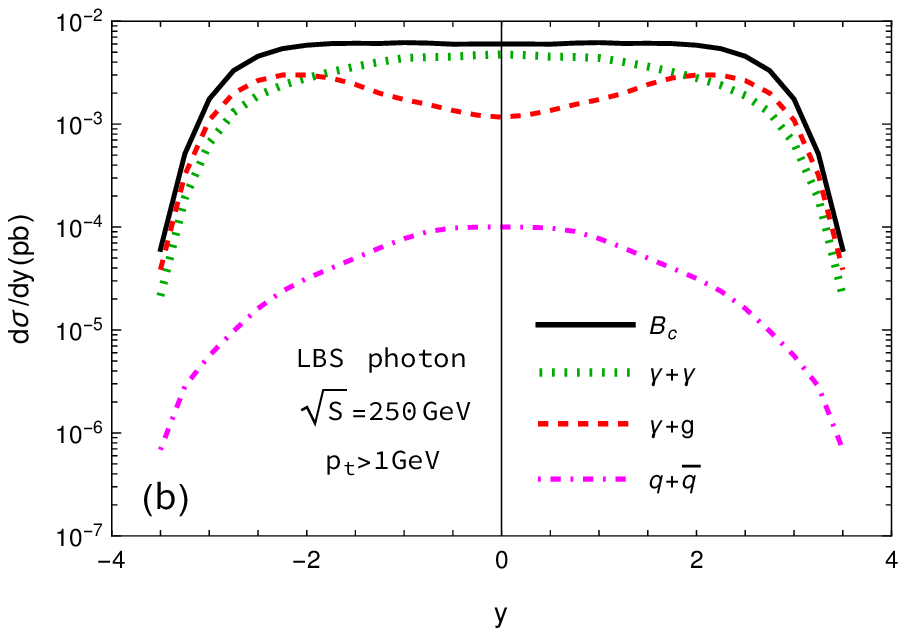}
	\includegraphics[width=.24\textwidth]{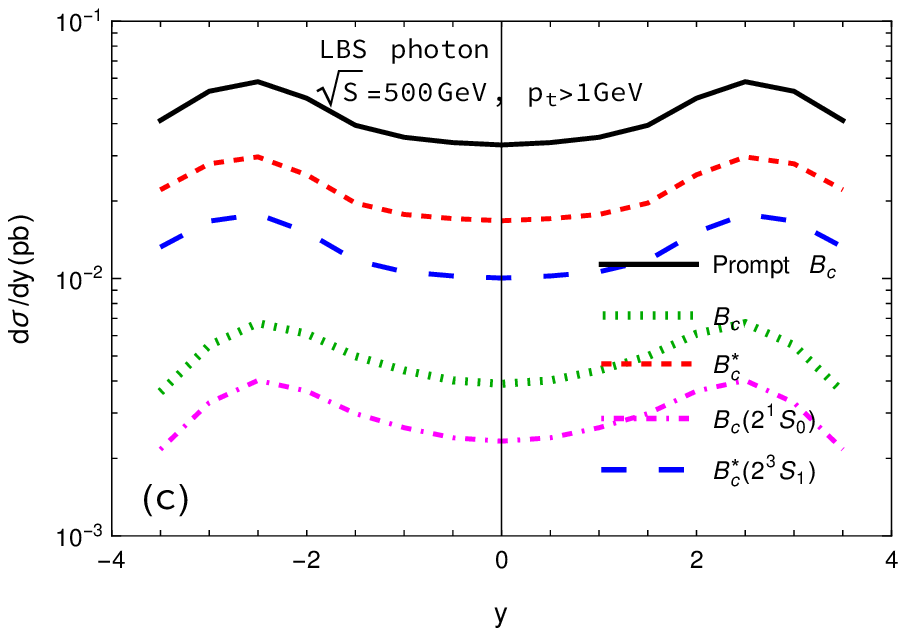}
	\includegraphics[width=.24\textwidth]{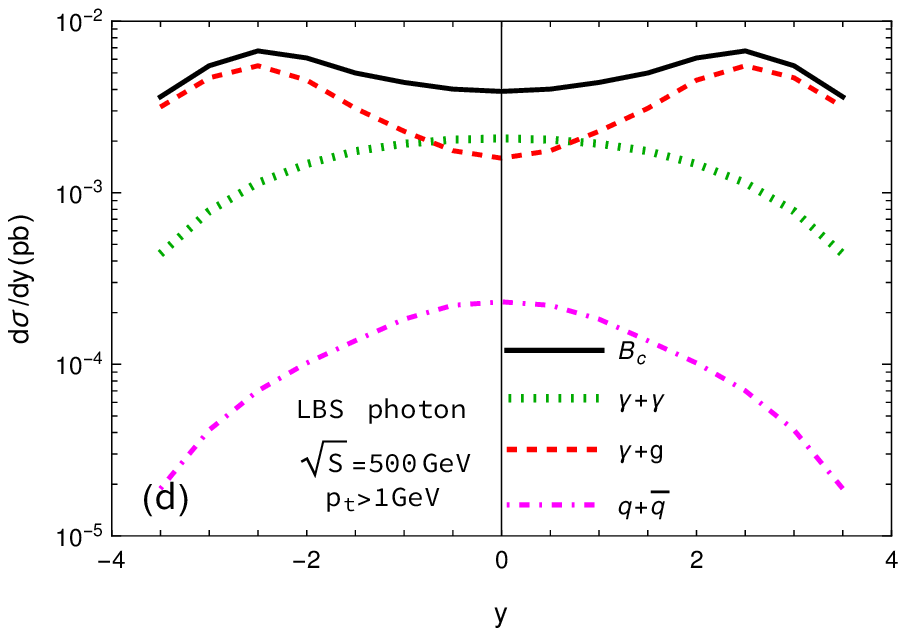}\\
	\includegraphics[width=.24\textwidth]{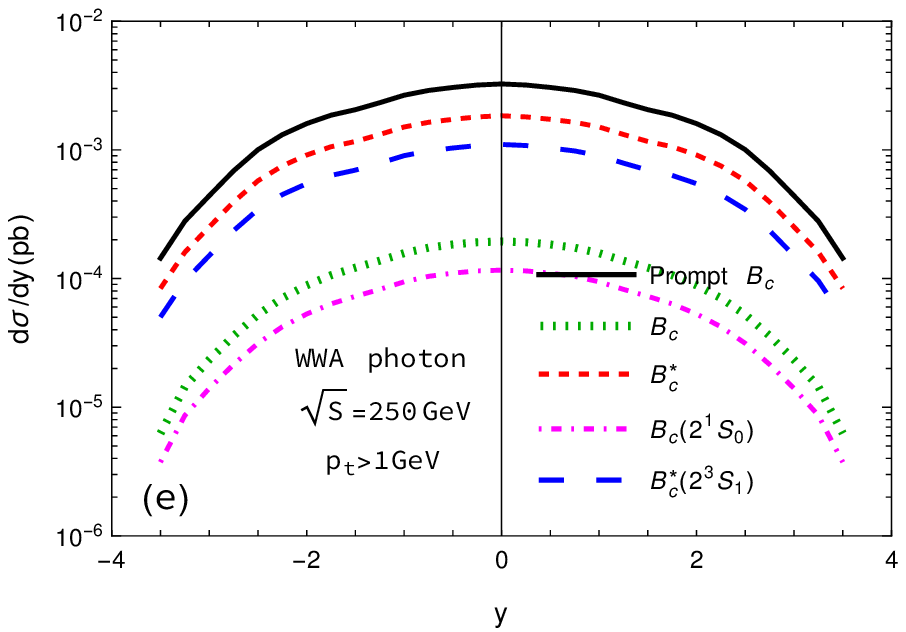}
	\includegraphics[width=.24\textwidth]{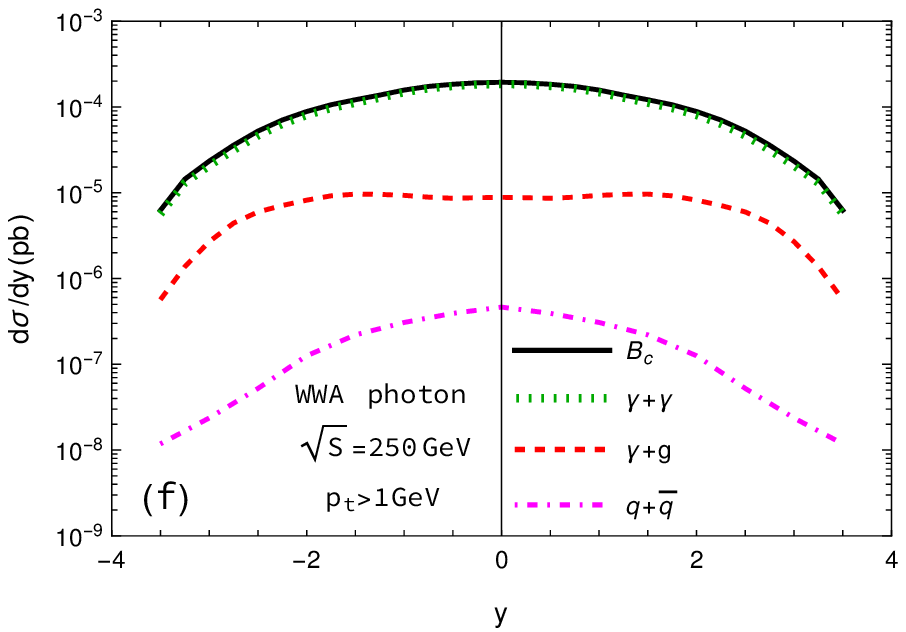}
	\includegraphics[width=.24\textwidth]{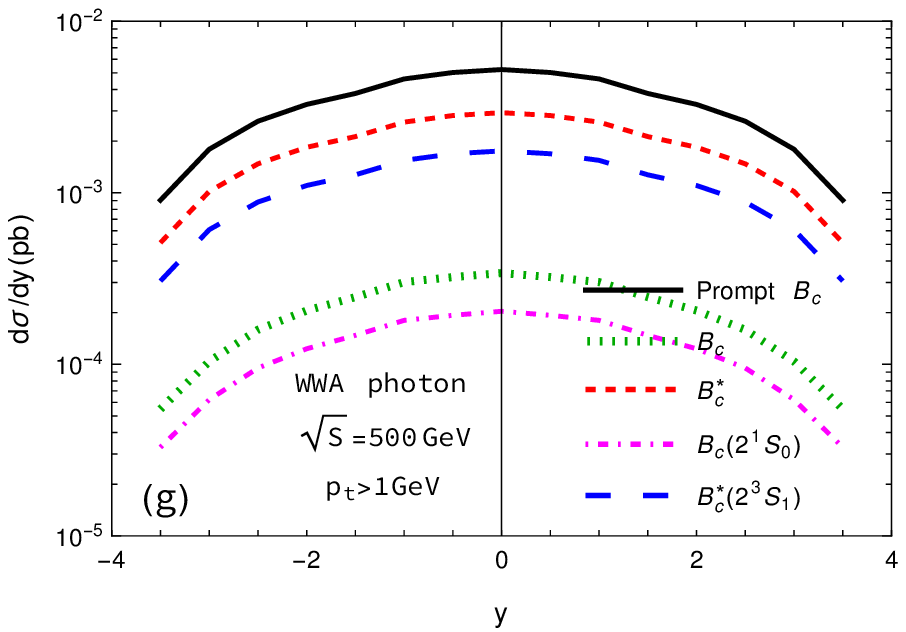}
	\includegraphics[width=.24\textwidth]{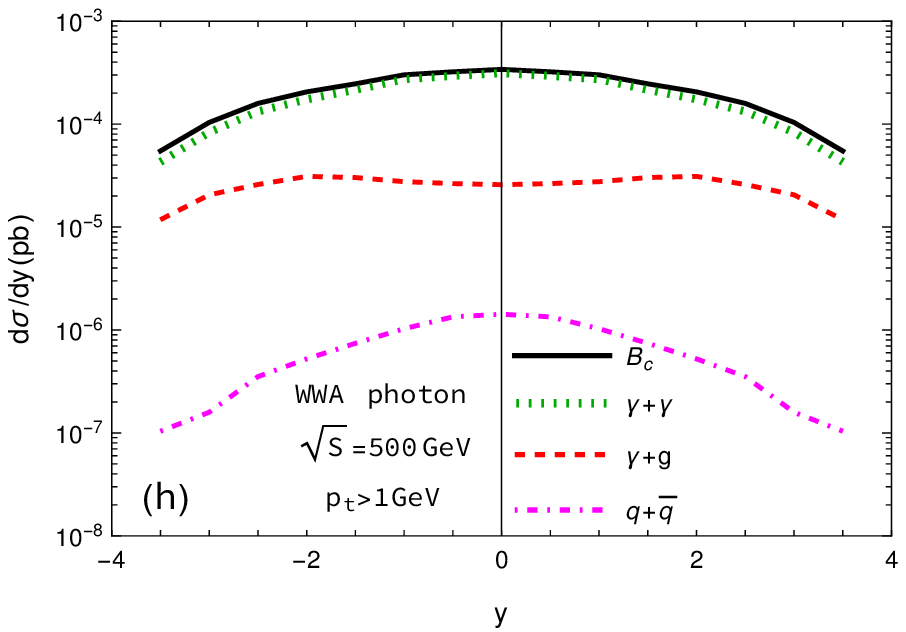}
	\caption{\label{fig:y} The rapidity ($y$) distributions for $B_c$ photoproduction. (a,c,e,g): $y$ distributions of four $B_c$ states and ``prompt $B_c$" means production of the ground $B_c$ after including the feed-down contributions from excited states with $100\%$ decay probability to it. 	(b,d,f,h): $y$ distributions for three channels in Eq.~(\ref{eq:channel-1},\ref{eq:channel-2},\ref{eq:channel-3}) of the ground $B_c$ production.}
\end{figure*}

The energy spectrum of the photon from the bremsstrahlung effect can be described in WWA, which takes the form~\cite{Frixione:1993yw},
\begin{eqnarray}
	f_{\gamma/e}(x) &=& \frac{\alpha}{2\pi}\Bigg[\frac{1 + (1 - x)^2}{x} {\rm log}\frac{Q^2_{\rm max}}{Q^2_{\rm min}} \nonumber\\
	&&+2m_e^2x\left(\frac{1}{Q^2_{\rm max}}
	-\frac{1}{Q^2_{\rm min}}\right)\Bigg],
\end{eqnarray}
where $x=E_{\gamma} / E_{e}$ is the fraction of the longitudinal momentum carried by the photon, $\alpha$ is the electromagnetic fine structure constant, $Q^2_{\rm min} = m_e^2 x^2/(1-x)$ and $Q^2_{\rm max} = (E\theta_c)^2(1-x) + Q^2_{\rm min}$. $\theta_c=32\mathrm{~mrad}$ is the maximum scattered angular cut in order to ensure the photon to be real. $E=E_e=\sqrt{s}/2$ with the collision energy $\sqrt{s}$.

And for the case of LBS photon, its spectrum function is~\cite{Ginzburg:1981vm},
\begin{eqnarray}
	f_{\gamma/e}(x)=\frac{1}{N}\left[1-x+\frac{1}{1-x}-4 r(1-r)\right],
\end{eqnarray}
where $r=x /\left[x_{m}(1-x)\right]$, and the normalization factor,
\begin{eqnarray}
	N&=&\left(1-\frac{4}{x_{m}}-\frac{8}{x_{m}^{2}}\right) \log(1+x_m)\nonumber\\
	&&+\frac{1}{2}+\frac{8}{x_{m}}-\frac{1}{2 (1+x_m)^{2}}.
\end{eqnarray}
Here $x_{m}=4 E_{e} E_{l} \cos ^{2} \frac{\theta}{2}$, $E_e$ and $E_l$ are the energies of incident electron and laser beams, respectively. $\theta$ is the angle between them. The energy of the LBS photon is restricted by
\begin{eqnarray}
	0 \leq x \leq \frac{x_{m}}{1+x_{m}},
\end{eqnarray}
with optimal value of $x_m$ being $4.83$~\cite{Telnov:1989sd}. These two spectra have quite different behaviors as shown in Fig.~\ref{fig:fre}.

Within the framework of NRQCD factorization, the cross section of $B_c$ photoproduction at the $e^+e^-$ collider can be factorized as~\footnote{For simplicity, here only formulas for the ground-state $B_c$ meson are presented and those for the excited states are similar.},
\begin{eqnarray}
		&&\mathrm{d} \sigma\left(e^{+} e^{-} \rightarrow e^{+} e^{-} B_c+b+\bar{c}\right)\nonumber\\
		 &&= \int \mathrm{d} x_{1} f_{\gamma / e}\left(x_{1}\right) \int \mathrm{d} x_{2} f_{\gamma / e}\left(x_{2}\right)\nonumber\\
		 && \times\sum_{i, j, k} \int \mathrm{d} x_{i} f_{i / \gamma}\left(x_{i}\right) \int \mathrm{d} x_{j} f_{j / \gamma}\left(x_{j}\right) \nonumber\\
		&&\times  \sum_{n} \mathrm{~d} \hat{\sigma}(i j \rightarrow c \bar{b}[n]+b+\bar{c})\left\langle \mathcal{O}^{B_c}[n]\right\rangle.
\end{eqnarray}
Here $f_{i/\gamma}$($i=\gamma,g,u,d,s$) represents the Gl\"uck-Reya-Schienbein (GRS) distribution function of parton $i$ in photon~\cite{Gluck:1999ub}.
$f_{\gamma/\gamma}(x)=\delta(1-x)$ is for the direct photoproduction process.
$d\hat{\sigma}(ij\to c\overline{b}[n]+b+\bar{c})$ is the differential partonic cross section, which can be calculated perturbatively. $c\overline{b}[n]$ is the intermediate state with quantum number $n$. $\langle{\cal O}^{B_c}[n]\rangle$ is the long distance matrix element(LDME) representing the probability for a $c\bar{b}[n]$ pair hadronizing into $B_c$ meson. In the lowest order approximation, only the color-singlet intermediate state $c\bar{b}[{}^1S^{[\textbf{1}]}_0]$ is considered and the corresponding LDME can be calculated by potential models. For definiteness, we consider the following sub-processes for three production channels,
\begin{eqnarray}
\label{eq:channel-1}
\gamma + \gamma \rightarrow B_c(B^*_c,B_c(2{}^1S_0),B^*_c(2{}^3S_1)) + b + \bar{c},\\
\label{eq:channel-2}
\gamma + g \rightarrow B_c(B^*_c,B_c(2{}^1S_0),B^*_c(2{}^3S_1)) + b + \bar{c},\\
\label{eq:channel-3}
q+\bar{q} \rightarrow B_c(B^*_c,B_c(2{}^1S_0),B^*_c(2{}^3S_1)) + b + \bar{c},
\end{eqnarray}
with $q=g,u,d,s$.
Some typical Feynman diagrams for calculating the partonic cross sections are shown in Fig.~\ref{fig:diag}.
In the analytical and numerical calculations, we use the well-established package Feynman Diagram Calculation (FDC)~\cite{Wang:2004du}, which employs the standard projection method~\cite{Bodwin:2002cfe} to deal with the amplitudes.

\section{\label{sec:3}Numerical results and discussions}

The input parameters in the calculation are taken as follows. The fine structure constant is fixed as $\alpha =1/137$. $m_b=4.8\mathrm{~GeV}$, $m_c=1.5\mathrm{~GeV}$ and $M_{B_c}=m_b+m_c$. The one-loop running strong coupling constant is employed. The renormalization scale is set to be the transverse mass of the $B_c$ meson, $\mu=\sqrt{M^2_{B_c}+p^2_t}$ with $p_t$ being its transverse momentum. The LDMEs $\left\langle \mathcal{O}^{B_c}[n]\right\rangle$ are related to the wave function at the origin, e.g., $\left\langle \mathcal{O}^{B_c}[n]\right\rangle \approx N_c|R_S(0)|^2/(2\pi)$, with $|R_{1S}(0)|^2=1.642\mathrm{~GeV}^3$ and $|R_{2S}(0)|^2=0.983\mathrm{~GeV}^3$~\cite{Eichten:1994gt,Eichten:1995ch}.

Table~\ref{tab:cross-section-1} lists the integrated cross section of the photoproduction of $B_c$ under three typical collision energies, both for the LBS photon and WWA photon. It is shown from the table that all cross sections become larger with the increment of collision energy. The cross sections via LBS photon are much larger than those of WWA photon. This is due to the quite different spectra functions of them as shown in Fig.~\ref{fig:fre}. We can also see that the production rate of the vector meson $B^*_c$ are much larger than those of the scalar $B_c$. Since the excited states decay to the ground state with almost $100\%$ probability, they will greatly increase the production of $B_c$. Taking as example the integrated luminosity of future $e^+e^-$ collider to be $\mathcal{O}(10^4)\mathrm{~fb^{-1}}$ and considering the feed-down contributions from excited $B_c$ states, we shall have about $3.3\times10^6$ ($1.3\times10^5$) $B_c$ mesons to be generated via LBS (WWA) photons under collision energy $\sqrt{s}=250\mathrm{~GeV}$. Thus the photoproduction of $B_c$ at future $e^+e^-$ colliders provides good opportunity to study $B_c$ meson.

Table~\ref{tab:cross-section-2} shows the contributions from different channels for the LBS photon. With the increase of $\sqrt{s}$, the cross section of direct photoproduction channel (Eq.~(\ref{eq:channel-1})) decreases, while those of the other two channels become larger. At $\sqrt{s}=250\mathrm{~GeV}$, the $\gamma+\gamma$ channel provides largest production and the single-resolved photoproduction $\gamma+g$ also gives significant contributions. With the increment of the collision energy, the channel $\gamma+g$ becomes dominant. Consequently for the LBS photon, the resolved photoproduction channels of $B_c$ at future $e^+e^-$ colliders should be taken into account.

The situations are quite different for the WWA photon, as shown in Table~\ref{tab:cross-section-3}. The cross sections of all the three channels become larger when increasing the collision energy. The $\gamma+\gamma$ channels are always dominant while contributions of other two channels are very small or even negligible.

Fig.~\ref{fig:pt} presents the transverse momentum distributions of $B_c$ photoproduction. All the distributions have a peak around several $\mathrm{~GeV}$ of $p_t$ and decrease logarithmically in the large region. From Fig.~\ref{fig:pt}(b,c), we can see that the single resolved channel $\gamma+g$ give important contributions in relative small $p_t$ region. In real experiments, there maybe not enough $B_c$ events in large $p_t$ region to make well measurements. Thus the single resolved photoproduction should be included in the calculation of $B_c$ photoproduction. For the WWA photon, the direct photoproduction channel $\gamma+\gamma$ is always primarily dominant in whole $p_t$ region.

Fig.~\ref{fig:y} presents the photoproduction in terms of the rapidity($y$) distributions of final $B_c$ mesons. There are wide plateaux within $|y|<2.5$ for the LBS photoproduction at $\sqrt{s}=250\mathrm{~GeV}$, while the curves carved in at $\sqrt{s}=500\mathrm{~GeV}$. This is because with the increase of collision energy, the contribution of $\gamma+g$ channel becomes dominant. The rapidity distribution of WWA photoproduction look ordinary compared with the LBS case.

\begin{table}
	\caption{\label{tab:uncer-mc}Variations of the integrated cross sections (in unit of fb) by $m_c$ for the photoproduction of $B_c$ via the LBS photon and WWA photon (in brackets) at $\sqrt{s}=250\mathrm{~GeV}$, respectively. The cut $p_t>1$ is imposed and three channels of Eq.~(\ref{eq:channel-1},\ref{eq:channel-2},\ref{eq:channel-3}) have been summed up.}
	\begin{ruledtabular}
		\begin{tabular}{cccc}
			$m_c(\mathrm{GeV})$ & $1.4$ & $1.5$ & $1.6$\\
			\colrule
			$\sigma_{B_c}$ & $41.09(0.92)$ & $33.50(0.73)$ & $27.64(0.59)$ \\
			$\sigma_{B^*_c}$ & $211.81(8.73)$ & $175.06(7.14)$ & $146.46(5.90)$  \\
			$\sigma_{B_c(2{}^1S_0)}$ & $24.60(0.55)$ & $20.05(0.44)$ & $16.55(0.35)$  \\
			$\sigma_{B^*_c(2{}^3S_1)}$ & $126.80(5.23)$ & $104.80(4.27)$ & $87.68(3.53)$ 
		\end{tabular}
	\end{ruledtabular}
\end{table}

\begin{table}
	\caption{\label{tab:uncer-mb}Variations of the integrated cross sections (in unit of fb) by $m_b$ for the photoproduction of $B_c$ via the LBS photon and the WWA photon (in brackets) at $\sqrt{s}=250\mathrm{~GeV}$, respectively. The cut $p_t>1$ is imposed and three channels of Eq.~(\ref{eq:channel-1},\ref{eq:channel-2},\ref{eq:channel-3}) have been summed up.}
	\begin{ruledtabular}
		\begin{tabular}{cccc}
			$m_b(\mathrm{GeV})$ & $4.6$ & $4.8$ & $5.0$\\
			\colrule
			$\sigma_{B_c}$ & $38.72(0.87)$ & $33.50(0.73)$ & $29.16(0.61)$ \\
			$\sigma_{B^*_c}$ & $201.84(8.59)$ & $175.06(7.14)$ & $152.76(5.97)$  \\
			$\sigma_{B_c(2{}^1S_0)}$ & $23.18(0.52)$ & $20.05(0.44)$ & $17.45(0.37)$  \\
			$\sigma_{B^*_c(2{}^3S_1)}$ & $120.83(5.14)$ & $104.80(4.27)$ & $91.45(3.57)$ 
		\end{tabular}
	\end{ruledtabular}
\end{table}

Take $\sqrt{s}=250\mathrm{~GeV}$ as example, we estimate theoretical uncertainties induced by the heavy quark masses and the renormalization scale. Table~\ref{tab:uncer-mc} presents the uncertainties from the variation of charm quark mass where we take $m_c=1.5\pm0.1\mathrm{~GeV}$ with $m_b=4.8\mathrm{~GeV}$ and $\mu=\sqrt{M^2_{B_c}+p^2_t}$. Table~\ref{tab:uncer-mb} shows the uncertainties from $m_b=4.8\pm0.2\mathrm{~GeV}$ with $m_c=1.5\mathrm{~GeV}$ and $\mu=\sqrt{M^2_{B_c}+p^2_t}$. Table~\ref{tab:uncer-mu} is for $\mu={\cal C}\sqrt{M^2_{B_c}+p^2_t} ({\cal C}=0.5,1,2)$ with $m_c=1.5\mathrm{~GeV}$ and $m_b=4.8\mathrm{~GeV}$. We see that slight variation of the heavy quark mass can result in substantial change of the cross sections. A strong dependence on the renormalization scale indicates it is maybe large that the next-to-leading order and higher orders of corrections in $\alpha_s$.

\begin{table}
	\caption{\label{tab:uncer-mu}Variations of the integrated cross sections (in unit of fb) by $\mu={\cal C}\sqrt{M^2_{B_c}+p^2_t}$ with ${\cal C}=0.5,1,2$, for the photoproduction of $B_c$ via the LBS photon and WWA photon (in brackets) at $\sqrt{s}=250\mathrm{~GeV}$, respectively. The cut $p_t>1$ is imposed and three channels of Eq.~(\ref{eq:channel-1},\ref{eq:channel-2},\ref{eq:channel-3}) have been summed up.}
	\begin{ruledtabular}
		\begin{tabular}{cccc}
			${\cal C}$ & $0.5$ & $1.0$ & $2.0$\\
			\colrule
			$\sigma_{B_c}$ & $45.69(1.07)$ & $33.50(0.73)$ & $25.77(0.53)$ \\
			$\sigma_{B^*_c}$ & $243.06(10.68)$ & $175.06(7.14)$ & $133.30(5.15)$  \\
			$\sigma_{B_c(2{}^1S_0)}$ & $27.35(0.64)$ & $20.05(0.44)$ & $15.43(0.32)$  \\
			$\sigma_{B^*_c(2{}^3S_1)}$ & $145.51(6.39)$ & $104.80(4.27)$ & $79.80(3.08)$ 
		\end{tabular}
	\end{ruledtabular}
\end{table}

\section{\label{sec:4}Summary}

In this work, we have investigated the photoproduction of $B_c, B_c^*, B_c(2{}^1S_0)$ and $B^*_c(2{}^3S_1)$ at future $e^+e^-$ colliders, where two sources of initial photons can be realized, i.e., the LBS photon and the WWA photon. Besides the direct photon-photon fusion, we specially consider the resolved photoproduction mechanisms that are generalized via the channels of Eq.~(\ref{eq:channel-2},\ref{eq:channel-3}), which are lack in previous studies. Numerical results show that the single resolved photoproduction channel (Eq.~(\ref{eq:channel-2})) can give sizable and even dominant contributions under certain collision configurations, e.g., the LBS photoproduction at $\sqrt{s}=500\mathrm{~GeV}$. Considering the excited states of $B_c$ could decay to the ground state with almost $100\%$ probability, they shall be important sources of $B_c$ production. If setting the integrated luminosity of future $e^+e^-$ collider to be $\mathcal{O}(10^4)\mathrm{~fb^{-1}}$ and considering the feed-down contributions from the excited $B_c$ states, we shall have about $3.3\times10^6$ ($1.3\times10^5$) $B_c$ meson events to be generated via the LBS (WWA) photons under the collision energy $\sqrt{s}=250\mathrm{~GeV}$. However, the leading-order calculations in $\alpha_s$ of the cross section have large dependence on the heavy quark masses and the renormalization scale, which means corrections of higher orders maybe substantial. We would like to leave them for future study. Nevertheless, we could expect that future $e^+e^-$ colliders to be good platforms for investigating the $B_c$ meson properties.

\begin{acknowledgments}

This work was supported in part by the Natural Science Foundation of China under Grants No. 12147116, No. 12175025, No. 12005028 and No. 12147102, by the China Postdoctoral Science Foundation under Grant No. 2021M693743 and by the graduate research and innovation foundation of Chongqing, China under Grant No.ydstd1912.

\end{acknowledgments}

%

\end{document}